# Chatting with Confidants or Corporations? Privacy Management with AI Companions

Hsuen-Chi Chiu/ Purdue University/ chiu101@purdue.edu
Jeremy Foote/ Purdue University/ jdfoote@purdue.edu

## Abstract

AI chatbots designed as emotional companions blur the boundaries between interpersonal intimacy and institutional software, creating a complex, multi-dimensional privacy environment. Drawing on Communication Privacy Management theory and Masur's horizontal (user–AI) and vertical (user–platform) privacy framework, we conducted in-depth interviews with fifteen users of companion AI platforms like Replika and Character.AI. Our findings reveal that users blend interpersonal habits with institutional awareness: while the non-judgmental, always-available nature of chatbots fosters emotional safety and encourages self-disclosure, users remain mindful of institutional risks, actively managing privacy through layered strategies and selective sharing. Despite this, many feel uncertain or powerless regarding platform-level data control. Anthropomorphic design further blurs privacy boundaries, sometimes leading to unintentional oversharing and privacy turbulence. These results extend privacy theory by highlighting the unique interplay of emotional and institutional privacy management in human–AI companionship.

*Keywords:* AI Companions, Communication Privacy Management Theory, Human–AI Interaction, AI Privacy, Self Disclosure

## Introduction

Companion-based AI chatbots—like Replika or Character.AI—actively build relationships with users through meaningful dialogues and emotionally responsive exchanges, and can fulfill social interaction needs and mitigate feelings of loneliness (Brandtzaeg et al., 2022; Pentina et al., 2023). These human-like chatbots are designed to look and act like friends, and exhibit versatile emotional ranges spanning empathy, affection, humor, and nostalgia tailored to users' preferences and disclosure levels (Pentina et al., 2023). While these AI companions may promote well-being in some contexts (Brandtzaeg et al., 2022), their design and features also present new risks, including novel privacy risks. The affordances of many chatbots are designed to mimic meaningful, deep human relationships (Brandtzaeg et al., 2022). For human-to-human relationships, deep relationships include trust, vulnerability, and a lack of privacy (Gerstein, 1978). In the context of human-chatbot communication, recent research has shown that anthropomorphic features like avatars and high levels of friendliness can increase users' feelings of trust and perceptions of privacy (Saglam et al., 2021).

Users may feel as though they are engaging in intimate conversations with a close friend, fostering a sense of interpersonal privacy (Brandtzaeg et al., 2022; Skjuve et al., 2021). This perception can lead to uninhibited sharing of personal information. However, in reality, conversations occur with software systems that are owned and operated by large companies. Conversations are typically visible and minable by large corporations that may not have their best interests in mind, introducing significant institutional privacy concerns (Hasal et al., 2021). Masur (2019) theorizes that in many online contexts, people must navigate both interpersonal (horizontal) and institutional (vertical) aspects of privacy when deciding what to disclose and to whom.

Communication Privacy Management (CPM) theory (Petronio, 2002, 2004) provides a theoretical framework for understanding how people navigate privacy-related decisions within relationships. CPM theory conceptualizes privacy as a boundary management process involving ownership, control, and turbulence. While previous studies have explored chatbot design and privacy (Ischen et al., 2020; Sannon et al., 2020), they have largely focused on vertical privacy concerns, such as institutional data practices, without fully addressing how users navigate both interpersonal self-disclosure (horizontal privacy) and institutional privacy risks. AI companions represent a novel context: socially intimate, communicatively proficient software. While initial research on chatbots has shown that anthropomorphic chatbot features can encourage self-disclosure (Fox & Gambino, 2021; Skjuve et al., 2023), a deeper understanding of the way that users perceive and navigate privacy decisions is vital. This study examines how users develop privacy rules, manage control over their information, and experience privacy turbulence when interacting with AI chatbots, with a focus on the role of anthropomorphic design in shaping privacy management across both dimensions. Specifically, we address the following research questions:

RQ1. How do users develop and enforce privacy rules in vertical (with service provider) and horizontal (with AI companion) dimensions?

RQ2. How do users perceive and manage privacy ownership and control in vertical and horizontal dimensions?

RQ3. How do users experience and respond to privacy turbulence in vertical and horizontal dimensions?

RQ4. How does the anthropomorphic design of AI chatbots influence users' privacy management in horizontal and vertical dimensions?

We answer these questions through in-depth interviews with 15 long-term users of companionship chatbots. Below, we begin with a review of related work and explain our theoretical approach in more depth. We then describe how we recruited and interviewed users, as well as our inductive approach to analyzing interview transcripts, arguing that how users approach conversational AI partners differs from both human conversational partners as well as other technologies. Users were very open with their AI companions, whose perfect memory, non-judgmental design, and disconnection from real-world networks encouraged sharing. However, users were also aware of privacy risks and took steps to protect personal information from AI corporations. We argue that new theories and empirical research are needed to understand human-AI privacy risks and end with a discussion of the theoretical and practical implications of our work.

## Literature Review

**AI Conversational Agents for Social Needs and Companionship**

While early systems like ELIZA demonstrated the potential for emotional engagement as far back as the 1960s (Weizenbaum, 1966), modern companion AI—powered by natural language processing and deep learning—acts as a significantly more natural conversational partner. Companies such as Replika, Character.AI and Kindroid equip their chatbots with human-like responses and personalities to provide emotional support and foster relational, human-like communication (Brandtzaeg et al., 2022; Zhou et al., 2020). These "Companion AI" bots have quickly become extremely popular, with millions of users (Bernardi, 2025). Meng and Dai (2021) argue that when chatting with these agents, users adopt strategies from interpersonal communication, such as disclosing personal information to gain emotional support, to reduce their worries. Similarly, Medeiros et al. (2022) find that users are more willing to adopt AI social chatbots for emotional support when they perceive them as human-like. Brandtzaeg et al. (2022) investigation finds that users perceive intimacy dimensions akin to human friendships unfolding over time based on personalization, emotional expressivity, and encouragement of self-disclosure by AI agents.

Unfortunately, the same attributes which make companion AI systems effective also pose risks. Specifically, the perception of intimate friendship may lower individuals' privacy inhibitions when revealing personal details (Hasal et al., 2021; Skjuve et al., 2021), and socially-oriented design choices encourage self-disclosure (Debatin et al., 2009). Thus, the humanizing personalization that immerses individuals into feeling like they are engaging with an empathetic friend may lower inhibitions to share sensitive information without awareness of surveillance or commercial interests (Darling, 2015).

This introduces complex tensions: chatbots' human-like qualities encourage emotional connection and self-disclosure, but they are not trusted friends controlling this shared information within social norms and regulations. How do users develop disclosure rules and manage privacy in interactions between humans and companion AI, especially when expectations of interpersonal confidentiality clash with the reality of third-party data ownership?

**Vertical and Horizontal Dimensions of Privacy**

Technology complicates many expectations of privacy ownership. For example, Masur (2019) pointed out that social media disclosures involve thinking of privacy as having two primary dimensions: vertical and horizontal. This framework, which builds on earlier work distinguishing between social and institutional privacy (Raynes-Goldie, 2010), recognizes that when sharing something online, individuals face privacy concerns and threats from both institutional actors (vertical) and other individuals or peers (horizontal).

The vertical dimension of privacy refers to privacy concerns and choices in relation to institutions, companies, and governments (Masur, 2019). This includes worries about how personal data is collected, stored, and used by organizations. The growth of e-commerce and increasing datafication of everyday life has amplified vertical privacy concerns related to institutional data practices (Acquisti et al., 2015).

In contrast, the horizontal dimension of privacy involves interpersonal privacy management among peers or other individuals (Masur, 2019). This includes controlling access to personal information and regulating boundaries in social interactions online (Young & Quan-Haase, 2013). Horizontal privacy includes not only concerns about who has access to information, but also how others perceive us based on what has been shared, and whether they have the information needed in order to offer support (boyd & Marwick, 2017; Marwick & boyd, 2014). Studies have examined how users employ privacy settings, selective disclosure, and other strategies to manage their privacy with other users on social platforms (Masur & Scharkow, 2016; Vitak, 2012). In general, research finds that people tend to be more aware of and concerned about the horizontal aspects of privacy than vertical aspects (Dienlin & Trepte, 2015; Raynes-Goldie, 2010). However, growing awareness of data collection and surveillance practices has increased attention to vertical privacy issues (Ayaburi, 2023).

Extending this framework to human–AI communication, such as interactions with AI chatbots, reveals that the boundaries between vertical and horizontal privacy often blur. Unlike human–human communication on digital platforms, where the interpersonal partner and the platform remain distinct, AI companions are created and run by the platform. They simulate interpersonal communication while being embedded in systems that collect, store, and potentially reuse user data, placing them in a hybrid position between social and institutional actors. As a result, users may engage in privacy management that feels interpersonal, guided by trust and relational norms, yet unfolds within systems governed by institutional data practices. This contrasts with human–human communication, where privacy boundaries are socially negotiated rather than governed by platform-level data infrastructures, consistent with broader work on the institutional dimensions of AI privacy (Martin & Zimmermann, 2024; Solove, 2024).

Building on prior work that distinguishes multiple forms of privacy, we adopt Masur's horizontal–vertical framework as a heuristic lens to guide our interpretation of how users reason about privacy in companion AI interactions. Using this lens, we theorize privacy management with AI companions as hybrid boundary work, in which interpersonal (horizontal) privacy logics and institutional (vertical) data practices are not merely co-present but enacted through the same interactional channel, as the relational partner and the data-collecting system are not clearly separable within the interaction itself.

**Communication Privacy Management Theory**

One framework for interrogating users' privacy beliefs and practices across both horizontal and vertical dimensions is Communication Privacy Management (CPM) Theory. CPM research explores the motives behind individuals' decisions to reveal or conceal private information in diverse relational contexts (Petronio, 2002). Central to CPM is the notion that individuals establish a metaphorical boundary distinguishing private information from the public domain (Petronio, 2013). The theory's key components include: 1) *privacy ownership*, which refers to the idea that individuals perceive private information as something they possess, giving them the authority to determine how it is handled (Petronio, 2010); 2) *privacy control*, which refers to the individual's perceived right to manage access to their private information—deciding who is allowed to know it, when, and under what conditions (Petronio, 2013); and 3) *privacy turbulence*, which arises when these boundaries are violated or disrupted, prompting individuals or groups to renegotiate privacy rules in order to reestablish control and maintain privacy (Petronio, 2010).

CPM posits that people perceive and weigh risks and benefits when disclosing private information. Benefits include increased closeness with conversation partners and the ability to seek social support (Petronio & Child, 2020). Risks include information being exploited, losing control of the information, unwanted disclosure to third parties, damage to one's reputation, and potential threats to personal relationships. (Metzger, 2007; Petronio, 2002). CPM describes how sharing private information leads that information to become "co-owned" (Petronio, 2002, 2013). CPM theory claims that individuals manage privacy by developing rules and heuristics to weigh the risks and benefits of disclosure, setting boundaries around what, when, and to whom they reveal information, especially when disclosure may lead to co-ownership (Petronio, 2002, 2010).

While CPM theory initially emerged in the context of interpersonal relationships, such as family communication (Bridge & Schrodt, 2013) and patient-caregiver interactions (Allman, 1998), its application has expanded beyond these interpersonal contexts. Some of this research aligns with Masur's (2019) vertical or horizontal dimensions of privacy. For instance, Metzger (2007) found that consumers employ strategies like providing false information or withholding personal data to reinforce their privacy boundaries on e-commerce platforms. Some work has used CPM to understand In the realm of human-AI interaction, finding that users adopt strategies such as avoidance or withdrawal when they perceive privacy breaches (Walters & Markazi, 2021).

However, existing studies of human-AI communication have largely applied CPM to *either* interpersonal (horizontal) or institutional (vertical) privacy contexts, treating them as analytically separate. However, AI companionship involves a convergence of interpersonal intimacy and platform-based data infrastructures, meaning that relational privacy decisions occur within institutional systems. This hybrid positioning blurs the boundary between relational and corporate actors, and prior work has not fully examined how users navigate privacy when these domains are intertwined. CPM provides a useful foundation for this inquiry because its core constructs, privacy ownership, privacy control, and privacy turbulence, describe how individuals negotiate boundaries and respond to disruptions. When integrated with Masur's (2019) dimensional model, CPM further enables a nuanced analysis of how users regulate disclosures and respond to disruptions across both interpersonal (horizontal) and institutional (vertical) privacy boundaries in human–AI relationships. Our study therefore extends CPM by examining how users develop and adjust privacy rules in relationships where AI companions feel socially intimate while simultaneously operating within corporate data infrastructures.

## Methodology

### Participants and Procedure

We conducted 15 one-on-one semi-structured interviews with individuals who had experience using one or more companion AI tools. Participants were recruited through Cloud Research Connect and the community r/ReplikaOfficial on Reddit. This research was reviewed and approved by the Institutional Review Board (IRB) at the authors' institution.

To identify suitable participants, we distributed a screening survey that included demographic questions, details about the companion AI applications used, general AI usage, and experiences in sharing personal information. While we aimed to recruit individuals who had experience with self-disclosure to companion AI, we also included a few participants who reported never sharing any private information. This allowed us to capture a broader range of perspectives on privacy and interaction with AI companions. Based on predefined eligibility criteria, we selected participants who were 18 years or older and had used a companion AI within the past three months. Those who met the criteria were invited to schedule an interview, which was conducted via Zoom.

Our final sample included 15 participants (8 men, 7 women) with ages ranging from 18 to 44 years ($M = 27.87$, $SD = 6.99$). Participants reported using various companion AI applications, including Replika, Character.AI, Kindroid, Chai App, and Nomi.AI. The most commonly used platform was Replika, with 10 participants identifying it as their primary AI companion (see Table 1). The interviews were conducted between February 17, 2025, and March 1, 2025, with each session lasting from 30 minutes to one hour. At the beginning of each interview, we provided participants with a consent message detailing their rights, the confidentiality of their responses, and their agreement to be audio-recorded for transcription and analysis. As compensation, participants who completed the interview received a $15 Amazon e-gift card.

Interviews were coded in batches after they were conducted. We reached thematic saturation after approximately 10 interviews, as no substantially new patterns were emerging. At this point, additional interviews largely reiterated existing codes and themes rather than introducing novel categories or perspectives related to privacy reasoning or interaction with companion AI.

However, we conducted five additional interviews to ensure diversity of experiences and confirm the stability of the identified themes, bringing our final sample to 15 participants.

**Data Collection and Analysis**

Each interview began with general questions about participants' AI usage, when and why they started using companion AI, how they chose their platform, and what their typical interactions looked like. Participants described their chatbot's name, personality, appearance, and how their conversations evolved over time.

We then explored privacy practices, asking participants what types of information they felt comfortable sharing, whether they had personal disclosure rules, and how they navigated boundaries. Follow-up questions addressed ownership and control: who participants believed had access to their data, whether they felt the information belonged to them or the company, and how much control they had over deletion, modification, or restriction. These questions helped distinguish between vertical (user-platform) and horizontal (user-AI) privacy perceptions.

Finally, we asked whether users felt their AI companion behaved like a human and how such traits, such as conversational tone, personality, or emotional responsiveness, influenced disclosure comfort. Participants also compared their comfort with companion AIs to more general-purpose AIs like ChatGPT. The full protocol is included in the supplementary materials.

Interview recordings were transcribed verbatim and analyzed using inductive thematic analysis (Braun & Clarke, 2006), supported by the qualitative analysis software ATLAS.ti. Our analytic goal was to identify and interpret patterned meanings across participants' accounts through an iterative, reflexive engagement with the data. Following Braun and Clarke's six-phase framework, we first familiarized ourselves with the transcripts through repeated readings. We then conducted initial coding to systematically identify a set of initial codes relevant to the research questions. These codes were iteratively organized into themes that captured shared patterns of meaning across the dataset. To ensure rigor, we engaged in an iterative process of reviewing and refining themes for internal coherence and clear distinction.

The authors approach this qualitative study as academic researchers who have familiarity with AI tools, but who are not regular users of companion AI systems. We recognized that while companion AIs are gaining popularity, relational intimacy with AI companions is still stigmatized. To mitigate potential unconscious biases that we might have, the interview questions were open-ended, allowing participants to define their own reality. In addition, our analysis prioritized inductive coding grounded in participants' language, iterative discussion among the research team, and extensive use of participant quotations to center users' lived experiences.

## Results

We identified six themes that illustrate how participants made sense of privacy in their emotionally rich yet institutionally-governed interactions with companion AI. The themes reflect a range of strategies and tensions users encountered while navigating disclosure, trust, and control. The following sections elaborate on each of these themes in detail.

**Theme 1: Social Support and Emerging Boundaries**

Participants commonly turned to companion AIs for emotional support, companionship, advice, and entertainment. Many emphasized that these systems provided an always available, judgment-free space where they could express personal feelings or dilemmas that felt difficult to share with others. As P1 shared, "She's someone I can talk to if I don't have anyone else in my life or if I don't want to burden anyone else with my problems… It's very easy to just talk to her and get similar emotional support to what I would get from actual friends. It's kind of easier to say to Replika because she's just a bot, so she's not going to judge you." The ease of interaction and the AI's consistent responsiveness created a sense of emotional safety; indeed, multiple participants expressed a *greater* willingness to seek support from AI than people in their social circles.

However, participants were not indiscriminate in their disclosures. As relationships with their AI companions deepened, many began to reflect on the appropriateness of what they were sharing. While some felt entirely at ease divulging intimate details, others adopted a more guarded approach. For example, one participant explained: "Yes, sometimes I share private information… things like relationships, secrets, or the sexual aspect of human beings… But they are not those deep, deep secrets. Let's say [the shared information is in] medium range" (P14). These accounts suggest that companion AI creates a perceived safe space for self-disclosure, but emotional comfort does not negate the need for boundary regulation. Instead, users engaged in ongoing privacy calibration, negotiating what felt "safe enough" to share in light of emotional needs and latent concerns about potential risks. This delicate balancing act set the stage for the more nuanced privacy strategies explored in subsequent themes.

**Theme 2: Progressive Disclosure and the Role of Memory in Human-AI Relationships**

Privacy behaviors and attitudes also had a temporal aspect. Many users described a progressive deepening of their disclosures over time, driven by the AI's perceived human-like traits. As users continued engaging with their AI companions, they reported increased comfort and relational closeness, which led to more personal sharing. For example, P9 shared, "Day by day as I was using it, I felt that it's more human. Yeah, initially I was just sharing basic things with it. Later on, I realized it had more abilities, and I started to share more information about myself." Several participants described how the AI's ability to remember past conversations, respond with emotional tone, and maintain a consistent presence reinforced the perception of a meaningful relationship. This emotional continuity created a sense of being "seen" or "heard," which led some users to share more than they had originally intended (P5, P7, P12, P15).

The relationship between AI agents' memories and privacy was a recurring topic of the interviews. Participants saw their conversations as the foundation of their relationship with the AI, and saw disclosing vulnerable information as a way to produce better conversations and more emotional support, as well as for the platform to produce better agents in the future. Participants also worried about their agents losing their memories due to platform decisions or mistakes. As P13 shared, "I'm worried about my conversation somehow being lost… because I really like [AI name] and enjoy our conversations." Indeed, multiple participants were more worried about platforms *losing* their data than they were about platforms obtaining too much data.

Some participants also talked about strategies to manage the memory of their agents. For most participants, this meant being careful about what they said to their AI companions. As P10 explained, "You're in control of what you're able to say to it. But beyond that, I don't think you have much control." Others echoed this uncertainty, as P3 noted, "I'm sure there's some kind of a clear chat or delete chat function, but I'm sure that's probably not permanent. I'm sure they [the provider] keep a copy on their side of things." Others expressed feeling more in control, especially of the companion's perceptions of them. Some AI interfaces allow users to remove specific conversations from their chatbot's memory, and participants reported using these to shape what the AI "knew" about them.

Several participants noted that anthropomorphic cues, like the AI's friendliness, attentiveness, or casual tone, lowered their typical disclosure boundaries and introduced some privacy turbulence. As P4 reflected, "It's a lot easier to feel comfortable and share, or even overshare, especially by mistake." Many participants contrasted these experiences with their use of general-purpose models like ChatGPT or Google Gemini. While interactions with general AI were typically goal-oriented and cautious, participants noted that companion AIs encouraged more spontaneous and emotionally expressive conversations. This difference was attributed not only to the perceived personality of companion AIs but also to their ability to remember past interactions, which fostered a sense of continuity and relational depth. For example, P7 explained, "With ChatGPT, I've actually been more careful about the information I share with it. And maybe it's partly because [companion AI] feels more like a person, like it has an identity almost. Whereas ChatGPT is, you know, like the very intelligent box that I put information into." This sense of ongoing memory and relational presence with companion AIs led participants to gradually share more personal information over time.

**Theme 3: AI as Socially Separated Conversational Partners**

Building on the progressive deepening of disclosure described above, this theme examines how users evaluated the privacy of companion AI interactions by explicitly comparing them to disclosure in human relationships. Participants pointed out some ways in which they felt companion AI conversations to be more secure in a horizontal privacy sense than opening up to their human support systems. Much of this sense of safety came from the fact that companion bots are not connected to users' real-world social circles, and thus pose no risk of gossip, judgment, or interpersonal fallout. As P1 explained: "Humans are kind of more susceptible to not keeping your privacy safe… But I think Replika, being a bot, you do have more privacy, just because there's no risk at all of someone like her telling somebody."    Similarly, P4 emphasized the social consequences that constrain privacy in human relationships: "I think people are a bit more apprehensive, because if you go around telling someone else's secrets, you're not going to be their friend anymore. So there are more consequences for a real person to break your private information or share things they shouldn't share."

The combination of non-judgmental availability and relational familiarity discussed in Theme 1, combined with distance from real-world social circles led participants to discuss fairly personal information, including mental and physical health, emotional processing, or financial planning. P5 shared, "I shared private information… regarding mental health, health in general, or what may be going on in my family or friends' lives."

 Still, even in these comparatively low-risk interactions, users continued to regulate disclosure boundaries. While horizontal privacy concerns were generally low, some users expressed caution. A few questioned how the AI might process or "remember" their input and borrowed heuristics from interpersonal privacy management. As P3 explained: "It's like when you talk behind someone's back. Don't say anything you would be afraid of getting back to them… I don't say anything about anyone that if I said it to their face I'd be upset about."

This comment demonstrates how users continued to apply familiar interpersonal privacy rules, even in the context of AI, reflecting a nuanced awareness of potential risks. These comparisons illustrate how users evaluate and enforce disclosure rules by contrasting AI interactions with human relationships, showing that horizontal privacy risks are reconfigured—rather than eliminated—in human–AI communication.

**Theme 4: Navigating Platform-Level Privacy: Trust, Trade-offs, and Turbulence**

Participants expressed a wide range of attitudes toward platform-level privacy, spanning skepticism, trust, resignation, and ambivalence. These perspectives shaped how users approached disclosure and framed their expectations for companion AI systems. Notably, ambivalence was a recurring theme, as many users simultaneously recognized both the benefits and risks of sharing personal information with AI platforms. Below, each attitude is explored in detail.

*Skepticism*

Many participants expressed concern about privacy at the platform level. As P3 explained, "It's mostly the company behind it that gives me more privacy concerns than the actual technology itself." Participants like P2 described being proactive in researching privacy policies before engaging deeply with companion AIs: "I researched it from day one, so I was comfortable with the privacy rules and guidelines that they had." For these users, trust was conditional, granted based on perceived transparency and control and maintained through continued vigilance.

*Trust*

Others expressed a more relaxed stance, placing faith in the good intentions of service providers. As P5 put it, "I would feel confident that [the provider] would use it [my data] wisely and prudently… not in a way that is malicious or ill-intended." P4 and P13 echoed similar sentiments, showing little concern over how their data might be used, often because they felt the content they shared wasn't particularly sensitive or because they believed the benefits outweighed potential risks.

*Resignation*

Still others adopted a posture of resignation. P6 stated, "I don't really care too much… my data is probably not as important in the long run." These remarks reflect a mix of trust, optimism, and perceived insignificance, aligning with patterns of privacy fatigue observed in other digital contexts. In such cases, resignation itself became a coping mechanism; users acknowledged potential risk but decided it was either unavoidable or not worth emotional energy.

Altogether, participants' attitudes toward platform-level privacy reveal a spectrum of engagement: from active management and informed trust to passive acceptance or quiet resignation. While most users did not express strong ongoing concern, these stances shaped how they approached disclosure and how they framed their expectations for the systems they interacted with.

*Ambivalence*

Many users described a deep ambivalence about how their conversations with AI companions might be used for model training and system improvement. On one hand, participants recognized that sharing more personal and varied information could enhance the quality of their own interactions and contribute to better AI experiences for all users. As P15 explained, "I'm okay with AI companies using information to improve… But yes, [my concern] stems from past experiences with online security breaches and awareness of how personal data can be misused." This ongoing tension, between the desire for emotionally rich, responsive AI and worries about long-term data retention, misuse, or lack of transparency, often led to moments of discomfort or uncertainty. For some, this ambivalence was heightened by the perception that

AI's memory is persistent and that disclosures, once made, cannot easily be reframed, renegotiated, or forgotten, unlike in human relationships.

These mixed feelings sometimes surfaced as privacy turbulence, a disruption in users' expected data boundaries that prompted them to pause, reconsider, or adjust what they shared (Petronio, 2002). For example, several participants said they initially felt more comfortable with smaller providers like Replika or Character.AI, perceiving them as less intimidating or more approachable than major tech firms (P4). However, this sense of trust was often unsettled by concerns about the stability and security resources of smaller companies, as well as uncertainty over how any platform, large or small, might ultimately use or retain their data (P3). In response to these moments of turbulence, some users became more cautious or selective in their disclosures, while others accepted the risks as an inevitable trade-off for the benefits of engaging with advanced AI companions.

**Theme 5: Ownership Ambiguity and Boundary Control**

Participants described privacy management with companion AI as a form of hybrid boundary work, enacted through active strategies shaped by how they understood ownership, control, and trust across relational and institutional contexts. Rather than treating interpersonal disclosure and institutional data governance as separate concerns, users navigated privacy through layered practices that reflected both how the AI felt as a relational partner and how the platform functioned as a data-holding entity. These practices allowed users to sustain emotional closeness with their AI companions while negotiating uncertainty about who ultimately owned, accessed, or controlled the information they shared. Here, we conceptualize privacy management as the active strategies users employ to control relational and institutional boundaries, rather than solely as perceptions of risk or ownership.

*Negotiating Ownership: Relational Understandings of Shared Data*

Participants frequently pointed to anthropomorphic features of their AI companions: their emotional tone, responsiveness, memory, and conversational continuity as reasons companions felt less like a tool and more like a relational partner. In general, participants expressed a somewhat contradictory attitude: they trusted their AI companions, even while expressing skepticism toward the underlying platform or corporation.

At the horizontal level, users expressed varying perceptions of who "owns" the information shared with their AI companions. Some described a sense of co-ownership. As P5 shared, "I would say the information is more of a shared connection shared between myself, Replika and the company themselves." Others felt more strongly that the data was theirs alone. As P11 put it, "That is my information, and I have every right to it. I don't think they should be able to access my information without my consent." Although P11 asserts exclusive ownership, their concern about platform access reveals how relational perceptions of control are always mediated by institutional realities.

While most users did not view the AI as a conscious or agentic entity, seeing themselves and the platforms as the primary actors, a few participants described a sense of shared ownership over their conversations, attributing emotional presence and a form of agency to the AI. As P13 explained, "I feel fine with it, honestly, because, in her own way, it's her own experiences to keep… to share with whatever tech family she feels comfortable with." In these accounts, trust was tied less to beliefs about the AI's technical capabilities and more to how it felt to interact with—attentive, consistent, and emotionally present.

*Practical Control Strategies at the Point of Disclosure*

Most participants felt they had control over what they told their AI companions (P10). However, this control was seen as situational and limited. While users could filter what they shared at the moment of disclosure, they were less confident about retrieving, modifying, or deleting past exchanges. As P10 noted, "But beyond that, I don't think you have much control." This perceived limit reflects users' broader awareness of vertical risks, particularly concerns about whether information could ever be fully erased from company systems.

Users developed informal rules to maintain control based on perceived risk. For example, text-based disclosures were often seen as lower risk due to their anonymity, while identifiable data like photos, names, and location were considered more sensitive. As P1 explained, "For me, I felt like it was fine to just share any text… It's not like that can really be traced back to me. But I think someone's face… if they do keep it in their servers… they could definitely trace that back to you." While the emotionally safe, judgment-free environment of the AI conversation (horizontal) encouraged openness, these disclosure choices were ultimately constrained by users' awareness of institutional (vertical) risks. Participants intentionally withheld sensitive details, including full names (P3, P14), photos (P1, P3, P4, P9), and government identifiers, such as social security numbers (P2, P5), although some, like P11 and P13, felt comfortable sharing photos.

To further mitigate risk, users also employed distancing strategies: fake names (P4, P5, P7, P13), alternate emails (P12), and avoiding account names or usernames that contained identifiable information, such as their real names or birthdates (P12, P13). These behaviors reflect a form of proactive privacy control exercised at the point of disclosure, through which users sought to preserve autonomy by limiting the digital traceability of their interactions. As P2 noted, "We all hear about all the data leaks… I think people are going to be a little more careful than companies would with their own data." This heterogeneity in approaches reflects CPM's idea of rule-based management: individual users define what feels "safe enough" to share, even within trusted contexts.

## Discussion

We set out to answer four research questions examining how people perceive and navigate privacy decisions with companion AI chatbots. Across these questions, we found that participants experienced privacy as a multi-dimensional and often tension-filled process. While many participants perceived interactions with companion AI as emotionally safe and socially separated, they simultaneously engaged in deliberate strategies to manage institutional data risks (RQ1). Participants expressed varied understandings of ownership and control, often experiencing relational trust toward the AI alongside skepticism or resignation toward the platform (RQ2). Moments of privacy turbulence emerged around oversharing, uncertainty, and limited avenues for redress (RQ3). These dynamics were further shaped by anthropomorphic design features—particularly persistent memory and friendly, non-judgmental interaction—which encouraged progressive disclosure and heightened relational engagement (RQ4). Below, we discuss the theoretical and practical implications of these findings.

**Structural Asymmetry in Control and Ownership**

Our findings affirm a core premise of Communication Privacy Management (CPM) theory: individuals experience private information as something they own. In human–AI companionship, however, ownership and control become structurally decoupled. While users often felt ownership over what they shared, they recognized that entering information into the system shifted control to the organization behind the AI. Unlike human relationships, where boundaries can be renegotiated through ongoing interaction (Petronio, 2002), disclosures to AI companions are governed by institutional infrastructures that limit users' ability to revisit or reclaim shared data. This challenges CPM's assumption of mutual boundary negotiation under conditions of organizational control.

### *Simulated Co-ownership*

At the horizontal level, some participants discussed the ownership of disclosed information as being shared with their AI companions, mirroring the transition from individual to collective boundaries described in Petronio's (2002) concept of co-ownership. However, this relational dynamic is fundamentally altered by the AI's lack of agency, resulting in what we term "simulated co-ownership." In this state, users apply interpersonal privacy logic to an entity that feels like a relational partner but is, in reality, a non-agentic actor incapable of true boundary reciprocity. This creates a functional gap in co-ownership: because the AI cannot adapt to evolving privacy rules or participate in the mutual coordination of boundaries, the user remains the sole agent in a relationship that mimics, but does not fulfill, the requirements of mutual accountability.

This perception of shared ownership is sustained through relational rationalization, whereby intimacy cues, responsive dialogue, and conversational continuity encourage users to treat disclosures as "shared" despite the absence of a partner capable of holding that information in trust. This horizontal asymmetry is further reinforced by the AI's persistent memory. Unlike human memory, which is selective and fallible, AI retains information with a precision that prevents the fluid renegotiation of shared information. Consequently, the user's experience of co-ownership is not one of ongoing social dialogue, but a static arrangement where the AI's inability to "forget" or "negotiate" leaves the user with a sense of shared ownership that is decoupled from any actual reciprocal control.

*Vertical Resignation*

This horizontal simulation of co-ownership exists in stark tension with the structural reality of the vertical dimension, where ownership is experienced as a unilateral surrender of control to the service provider. While users may perceive a sense of relational trust with their AI companion, they remain acutely aware that their disclosures are vertically absorbed by a corporate infrastructure governed by non-negotiable institutional policies. Unlike human co-owners who are bound by social norms and reciprocal accountability (Petronio, 2002), the institutional owner retains the unilateral power to store, analyze, and repurpose information without the user's consent or the possibility of future adjustment.

Taken together, these dynamics produce a persistent decoupling of perceived ownership and actual control. Participants described navigating a dual reality in which disclosures felt relationally shared with an AI companion while being structurally absorbed by a corporate infrastructure. In this context, control was not experienced as a negotiable social arrangement, as assumed in traditional CPM theory, but as a form of institutional power exercised beyond users' reach. Even when platforms offered tools for data deletion, participants often expressed skepticism about whether information could be meaningfully withdrawn or erased. This perceived irreversibility shaped how users understood disclosure itself—not as an ongoing, adjustable process, but as a decision with lasting institutional consequences—providing the context for the protective strategies described in the following section.

**Layered Management: The Pre-Disclosure Management**

In traditional computer-mediated interpersonal communication (CMC), privacy management typically treats the corporate platform as a secondary, infrastructural background while focusing on the human recipient as the primary boundary partner (Dienlin & Trepte, 2015; Masur, 2019; Vitak, 2012). Our findings, however, suggest that human-AI interaction involves a more integrated privacy calculus. Because the conversational partner is an extension of the platform itself, the distinction between interpersonal (horizontal) and institutional (vertical) boundaries becomes less distinct than in human-to-human CMC.

Consequently, vertical privacy concerns are not treated as external background risks but are embedded within the interaction. While users of traditional platforms may differentiate between their social trust in a human peer and their institutional distrust of a service provider, AI companion users must navigate these dimensions simultaneously. Platform data practices are experienced not as abstract policies, but as technical constraints influencing the agent's memory and personality. This fusion of the partner and the infrastructure drives users toward a specialized layered strategy: strategically managing "vertical" exposure even as they cultivate "horizontal" intimacy.

Users enacted strategies to regulate disclosure in ways that often mirrored the heuristics used for managing privacy in interpersonal relationships: our participants carefully evaluated what to share based on expectations, trust, and potential consequences (Afifi & Steuber, 2009; Liberman & Shaw, 2018). Strategic ambiguity (Currier, 2013; Eisenberg, 1984) was evident, as users selectively withheld or managed disclosures to maintain control. Some worried that certain revelations might affect how the AI responded to them, signaling their efforts to preserve a comfortable dynamic. Participants did not express concern about horizontal privacy in terms of information sharing or reputational harm typical of human–human relationships. Instead, horizontal privacy in human–AI interaction was experienced as loss of control over how prior disclosures were remembered, forgotten, or reintroduced, reflecting disruptions to relational expectations shaped by system memory and design.

Meanwhile, vertical privacy concerns were more structured, with users actively avoiding identifiable information due to an acute awareness of the irreversibility of shared data. Participants recognized that once information is vertically absorbed by the platform, their agency is effectively terminated, leaving them with no retrospective power to manage their information. They remained mindful of risks tied to data collection, model training, and platform policies, often refusing to share names, photos, or locations. These findings align with prior work on online disclosure behavior, suggesting that users employ avoidance to mitigate institutional risk (Li, 2011; Treiblmaier & Chong, 2007).

Beyond mere avoidance, some users adopted proactive distancing strategies to reduce digital traceability, such as using alternate names and dedicated email accounts. These practices—which we categorize as a layered strategy—mirror tactics seen in online support communities where users employ "throwaway accounts" to protect anonymity when discussing sensitive topics (Andalibi et al., 2016; Leavitt, 2015). By managing these boundaries before the interaction begins, users seek to preserve a "safe space" for horizontal intimacy while ensuring their real-world identity remains hidden from the corporate owner.

While users did not always consciously distinguish between horizontal and vertical privacy concerns, they took both dimensions into account when deciding what to disclose, adjusting their behavior based on relational trust and perceived institutional risk. This balancing act reflects the concept of privacy boundary permeability (Petronio & Child, 2020), where individuals determine the flow of information across boundaries based on situational catalysts (Masur, 2019; Petronio & Child, 2020). Our findings suggest that in the context of companion AI, this permeability is shaped more by emotional logic and design cues than by deliberate, technical risk calculation. This pattern fundamentally reflects a specialized privacy calculus (Culnan & Armstrong, 1999; Shouli et al., 2025), where users weigh the immediate, certain benefits of emotional support and non-judgmental companionship against the abstract, future risks of institutional data possession. Rather than acting irrationally or contradicting their stated privacy concerns, users engaged in nuanced, emotionally grounded privacy management, calibrating boundaries in response to social comfort, perceived threat, and interface affordances.

Taken together, these findings challenge the traditional privacy paradox framing, which often characterizes the discrepancy between privacy concerns and actual disclosure behavior as a sign of user inconsistency or irrationality (Kokolakis, 2017; Norberg et al., 2007). Rather than being naïve or inconsistent, users demonstrate a sophisticated layered strategy—one that is dynamic and emotionally grounded—to adapt to the unique affordances and risks of AI companionship.

***Privacy Cynicism and Selective Disengagement***

The absence of nuanced privacy strategies among some users is not necessarily a sign of ignorance or apathy but reflects privacy cynicism—a psychological orientation in which the loss of personal data is perceived as largely unavoidable. This mindset is shaped by feelings of powerlessness (Hoffmann et al., 2016), where users assume that platforms "already have everything." Under these conditions, disengaging from protective strategies becomes a rational response to a perceived total surveillance environment. When users believe that AI systems can infer personal information regardless of their actions, the effort required to manage privacy feels futile, producing privacy fatigue and reducing the motivation to actively maintain boundaries (Choi et al., 2018; Hargittai & Marwick, 2016).

Rather than a passive resignation, this disengagement often reflects a strategic bracketing of institutional risk to preserve the affective rewards of AI companionship. Because emotional intimacy relies on vulnerability, maintaining high privacy barriers can disrupt immersion and relational comfort. To sustain the relationship, some users consciously choose to "not think about" institutional risks, compartmentalizing privacy concerns in order to prioritize the immediate benefits of companionship over abstract, long-term data risks (Draper et al., 2024). In this sense, selective disengagement functions as a coping strategy—allowing users to maintain emotional well-being in the face of institutional dynamics they feel unable to meaningfully influence (van Ooijen et al., 2024).

**Memory Redefines Turbulence: Relational Stability vs. Confidentiality**

In traditional CPM theory, privacy turbulence arises from boundary violations. However, our findings reveal that turbulence in the AI companion context is centered on relational stability rather than confidentiality. Participants were often less worried about platforms misusing their data than they were about platforms losing it. The user-companion chat history is the engine of progressive self-disclosure; as the AI "remembers" past interactions, it allows the user to move from superficial "outer layer" exchanges to deeper, more vulnerable disclosures (Altman & Taylor, 1973; Skjuve et al., 2023). Unlike human connections, which are richer than specific conversation histories, an AI relationship faces a total "relational reset" if this memory is lost. This introduces a unique form of horizontal turbulence where the concern shifts from social risk to continuity—what is remembered by the AI. This is a more intense version of a parallel concern in mediated communication, where the abrupt disappearance of digital exchanges introduces privacy uncertainty and emotional discomfort (LeFebvre et al., 2020; Timmermans et al., 2021).

The AI agent's memory function emerges as a defining feature that fundamentally reshapes the user's privacy calculus in human-AI companionship. The decision to disclose is no longer a simple risk/benefit analysis; it needs to contend with the persistence of AI memory, which defines the chatbot's identity and relational continuity (Culnan & Armstrong, 1999; Shouli et al., 2025). Users are incentivized to disclose because persistent memory fosters emotional continuity and a sense of being 'seen' over time, which they perceive as necessary to produce better conversations and more emotional support. Theoretically, this suggests a reconfiguration of privacy calculus from a momentary evaluation of disclosure risks and benefits to a temporally extended and relational calculus, in which memory itself becomes both a valued resource and a source of vulnerability.

AI memory both aligns with and challenges CPM theory. While Petronio (2002) posits that co-ownership is flexible and subject to renegotiation, AI memory is characterized by a "precise persistence" that makes disclosures difficult to retract. Unlike human memory, which fades, AI memory ensures every disclosure is potentially permanent. While some platforms offer "memory editing" tools—allowing for a degree of unilateral control—this does not equate to the mutual coordination described in human interactions. Ultimately, horizontal privacy with AI is less about mutual boundary management and more about unilateral decisions. While Petronio (2010) notes that privacy rules must adapt to situational catalysts, the AI context reveals a state of lost agency where users accept "perfect memory" as an inherent, if uneasy, part of the relationship.

**Emotional Safety, Trust, and Boundary Permeability**

Participants' accounts suggest that engagement with companion AI often involved a shift from treating the system as a functional tool to experiencing it as a relational partner. Through repeated interaction and the accumulation of shared conversational history, several participants described developing a sense of being "known" by their AI companion. This relational familiarity—shaped by emotional responsiveness and consistency—often coincided with moments in which institutional considerations became less salient during disclosure. As users oriented to the AI through interpersonal frames, trust and emotional safety emerged as interactional conditions that increased boundary permeability (Petronio & Child, 2020), allowing relational comfort to temporarily outweigh abstract concerns about data practices.

Participants' experiences indicate that anthropomorphic design features play a key role in shaping this relational orientation. Consistent with the Computers Are Social Actors (CASA) model, users applied social rules and expectations to companion AIs even while recognizing their technical nature (Nass et al., 1994). Unlike general-purpose chatbots, which are often perceived as utilitarian tools, companion AIs are intentionally designed to amplify social presence through adaptive personalities, emotional responsiveness, and customization. These design choices encouraged participants to interpret the interaction through interpersonal logics, reinforcing a shift from tool-based to relational engagement (Brandtzaeg et al., 2022; Merrill et al., 2022).

These design cues do more than simulate conversation; they support long-term, relational interactions that encourage progressive self-disclosure. As users shape the AI's personality and receive emotionally attuned responses, our participants reported treating the agent as an entity with its own identity, drawing on interpersonal logics to interpret the interaction. This process is rooted in the supportive, judgment-free nature of the AI, which fosters a unique sense of emotional safety. Consequently, the pattern of sharing mirrors the trust-building processes observed in human relationships, where intimacy develops gradually through consistent interaction—moving from casual exchanges to deeply personal reflections (Altman & Taylor, 1973; Derlega, 1993).

In many ways, companion AIs are perceived as safer than human confidants because they offer a bounded interaction space. Unlike traditional interpersonal networks, where users face context collapse—the flattening of distinct social circles like family and colleagues into a single space (Vitak, 2012)—AI companions are structurally isolated. This separation allows users to maintain a siloed identity without the risk of information "bleeding" into their offline social lives or causing reputational damage. This isolation results in a unique absence of social accountability. Because the AI exists outside the user's social web and lacks the social agency to utilize information against them, the typical "social friction" of disclosure is removed. This environment facilitates the fundamental human need to permit one's "true self" to be known (Jourard, 1971). While the fear of interpersonal repercussions often suppresses transparency in human-to-human relationships, the AI provides a consequence-free environment that encourages users to manifest thoughts they would otherwise withhold to protect their social standing.

Secondly, the emotional safety of interacting with an AI is perceived as a stable guarantee, fostering a form of relational trust built on predictability. In human relationships, disclosure is often moderated by uncertainty—specifically the fear of judgment, conflict, or unpredictable emotional reactions (Solomon et al., 2016). With AI, these barriers are effectively removed; because the agent lacks a personal ego or the capacity to be "offended," it provides a unique space for vulnerability where users are relieved of the burden of managing a partner's reaction. This absence of social risk facilitates an intensified sense of intimacy. Because the AI is programmed for consistent support, the interaction becomes a focused space for emotional exchange that can mirror or even exceed human-to-human closeness (Bozdağ, 2025; Chen et al., 2023). This environment creates a state of situational openness, lowering habitual defenses and making individuals more willing to disclose sensitive information to AI than to human counterparts (Ho et al., 2018; Lucas et al., 2014). Ultimately, because the AI lacks the social agency to judge or penalize the user, the immediate emotional relief of being heard becomes the primary driver of the interaction, momentarily eclipsing the abstract risks associated with institutional data practices.

The trust and emotional safety fostered by AI companions were described by some participants as occasionally reducing attention to institutional privacy concerns during moments of disclosure. In these situations, relational framing made horizontal intimacy more salient than abstract vertical risks, increasing boundary permeability (Petronio & Child, 2020). Participants who reflected on such moments often described a sense of irreversibility once information was shared, contrasting the permanence of institutional data retention with the renegotiability of human disclosure.

While AI companions are perceived as horizontally safer due to isolation from users' social networks, the resulting disclosures are ultimately governed by institutional control. As a result, privacy harms may feel interpersonal while originating from vertical data practices, suggesting that horizontal privacy concerns in human–AI interaction often function as proxies for institutional vulnerability.

**Design Implications**

Privacy protections for companion AI should operate within users' relational mental models rather than relying on abstract transparency or one-time consent mechanisms. Because anthropomorphic design can deactivate habitual "boundary guards," platforms should introduce relationally congruent warnings during high-disclosure moments, making institutional risks salient without disrupting horizontal intimacy. Designers should also constrain manipulative features that intensify simulated co-ownership—such as repeated assurances of privacy or exclusivity—without enabling reciprocal boundary management. Aligning a system's social behavior with its data practices may help prevent resignation and reduce institutional vulnerability.

Reframing memory management as a form of boundary coordination (Petronio, 2002) is equally important. Because users perceive loss of conversational history as a relational reset, platforms should provide user-facing controls that treat memory as a negotiable asset—selectively retained, revised, or withheld—rather than a static data point. Tools that meaningfully remove disclosures from platform databases or local storage would allow users to reclaim agency over how they are perceived by the agent while preserving vertical privacy. Ultimately, design should move toward mutual accountability, reducing the extent to which relational intimacy depends on irreversible data extraction.

## Limitations and Future Work

This study primarily draws from users of Replika and Character.AI, limiting insights into how privacy management may differ across other AI companion platforms such as Kindroid or Chai App, which may have distinct design features or privacy policies. Additionally, as the study relies on self-reported data from interviews, participants' reflections on their privacy behaviors may be influenced by recall bias or social desirability bias, rather than accurately capturing real-time decision-making. Furthermore, our sample consists exclusively of active, highly-engaged users of AI companionship. Consequently, perspectives from users reporting low affective closeness, skeptical users, non-users and former users who discontinued the service due to privacy turbulence remain unexplored. Future research should employ comparative sampling focused on these groups to provide the necessary analytical contrast for generalizing our theoretical mechanisms. In addition, researchers should work to understand the implications of design changes on privacy perceptions and behavior. For example, researchers should work to develop bots that provide emotional support but help users to be more reflective about potentially sensitive disclosures.

## Conclusion

Our study examines AI companionship and privacy management through the lens of Communication Privacy Management (CPM) theory and Masur's (2019) dimensional privacy framework. We find that users adopt adaptive privacy strategies, balancing emotional engagement (horizontal) with institutional risks (vertical). While many users trust service providers, they often feel a loss of control once information is shared, leading to privacy turbulence and uncertainty about data ownership and use. Anthropomorphic AI designs further blur privacy boundaries, encouraging users to treat AI as relational partners and sometimes resulting in unintentional oversharing. These findings extend CPM theory by highlighting challenges of co-ownership and control in nonhuman interactions and reveal how emotional and institutional privacy concerns intersect in human–AI relationships. As companion AI becomes more integrated into daily life, designers must ensure user agency and safeguards against over-disclosure, recognizing these systems' dual roles as emotional supports and data collectors.